\documentclass[12pt]{article}
\usepackage{graphicx}
\usepackage{type1cm, amsmath, amssymb}

\newcommand{\nn}{\nonumber}
\newcommand{\Zb}{\mathbb{Z}}

\newcommand{\Cb}{\mathbb{C}}

\newcommand{\del}{\partial}

\newtheorem{theorem}{Theorem}
\newtheorem{proposition}{Proposition}

\newcommand{\vp}{\varphi}
\newcommand{\Ct}{\widetilde{C}}

\newcommand{\delb}{\bar{\del}}
\newcommand{\psib}{\bar{\psi}}
\newcommand{\bC}{\overline{C}}
\newcommand{\spe}{\text{A-model}}
\newcommand{\Dcal}{{\cal D}}
\newcommand{\deldel}[2]{\frac{\del #1}{\del #2}}
\newcommand{\Zt}{\widetilde{Z}}
\numberwithin{equation}{section}

\begin{document}

\thispagestyle{empty}

\begin{flushright}
 \begin{tabular}{l}
 {\tt hep-th/0406078}\\
 \end{tabular}
\end{flushright}

 \vfill
 \begin{center}
{\LARGE
 \centerline{ Topological String Partition Functions as Polynomials} 
 \vskip 3mm
 \centerline{ }
}

 \vskip 2.0 truecm
\noindent{ \large  Satoshi Yamaguchi and Shing-Tung Yau} \\
{\sf yamaguch@math.harvard.edu~,~
 yau@math.harvard.edu}
\bigskip

 \vskip .6 truecm
 {
 {\it 
Harvard University, 
Department of Mathematics,\\ 
One Oxford Street,
Cambridge, MA 02138, U.S.A. 
} 
 }
 \vskip .4 truecm

 \end{center}

 \vfill
\vskip 0.5 truecm

\begin{abstract}

We investigate the structure of the higher genus topological string amplitudes
on the quintic hypersurface. It is shown that the partition functions of the 
higher genus than one can be expressed as polynomials of five generators. 
We also compute the explicit polynomial forms of the partition functions
for genus 2, 3, and 4. Moreover, some coefficients are written down for all 
genus. 
\end{abstract}
\vfill
\vskip 0.5 truecm

\newpage
\setcounter{page}{1}

\section{Introduction}

The topological string theory on a Calabi-Yau 3-fold is a good toy model
as well as a computational tool of the superstring. Especially, it is a
practicable problem to calculate the higher genus amplitudes,
while those of the physical string are technically difficult to calculate.
The topological 
A-model partition functions on a Calabi-Yau 3-fold $M$ is defined for
genus $g \ge 2$ as
\begin{align}
 F^{\spe}_{g}(t)=\sum_{d\in H_{2}(M,\Zb)}
   e^{- d \cdot t}\int_{\overline{\cal M}_{g,d}}1,
\end{align}
where $\overline{\cal M}_{g,d}$ is the compactified moduli space of degree $d$
holomorphic map from genus $g$ Riemann surface to $M$, and $t$ is a vector of 
complexified K\"ahler parameters. In the physical view 
point, the partition function can be defined similar way to the bosonic string 
as follows. Let us consider the
A-twisted theory of the sigma model on the Calabi-Yau $M$. It becomes a 
$N=2$ topological CFT which depends on the K\"ahler parameter $t$. We denote 
by ${\cal M}_{g}$ the moduli space of the complex structure on genus $g$ Riemann
surface. Then the genus $g\ge 2$ topological string partition function can be 
defined as
\begin{align}
 F_{g}(t,\bar t)=\int_{{\cal M}_{g}}\left\langle \prod_{k=1}^{3g-3}
\left(\int G^{-}\mu_{k}\right)
\left(\int \bar G^{-}\bar \mu_{k}\right)
\right \rangle_{g,t},
\end{align}
where $\mu_{k}$'s are the Beltrami differentials and $G^{-},\bar G^{-}$
are ``$b$-ghosts'' of $N=2$ topological CFT. The two definition are connected by
a gauge transformation $f(t)$ and the limit, namely
\begin{align}
 F_{g}^{\spe}(t)=\lim_{\bar t\to \infty}f(t)^{2-2g}F_{g}(t,\bar t).
\end{align}
The non-holomorphic partition function $F_{g}(t,\bar t)$ has a good global 
properties.

The $\bar{t}$ dependence is governed by the holomorphic anomaly equation which
is written down by Bershadsky, Cecotti, Ooguri and Vafa
\cite{Bershadsky:1993ta,Bershadsky:1994cx}. This equation provides an effective method to calculate the higher genus amplitudes. But there remain
some ambiguities to determine the amplitudes by using the holomorphic anomaly 
equation. Ref. \cite{Bershadsky:1994cx} have used geometric consideration to
fix the ambiguity, and obtain genus 2 partition function for the 
quintic hypersurface. As Ghoshal and Vafa have pointed out in 
\cite{Ghoshal:1995wm}, 
comparing the conifold limit and the topological string on conifold
gives non-trivial information.
 In \cite{Katz:1999xq}, Katz, Klemm, and Vafa have used the M-theory picture 
and obtained genus 3 and 4 partition function for the quintic.
In order to proceed this calculation, we want to 
understand the structure of the higher genus amplitudes.

Higher genus amplitudes are expected to have the property similar to the modular
forms. Every modular form can be written in a quasi-homogeneous polynomial of
Eisenstein series $E_4$ and $E_6$. This is the very beginning of the interesting
theory of the modular forms. It will be interesting as well as useful if
the topological string partition function has this kind of polynomial structure.
Actually, as pointed out in \cite{Candelas:1991rm}, a discrete group similar to 
$SL(2,Z)$ but not the same, act to the moduli space of the quintic hypersurface. 

In this paper, we explore the structure of the higher genus amplitudes of the quintic hypersurface.
We will show that the topological string partition function
$F_{g}$ can be written as a degree $(3g-3)$ quasi-homogeneous 
polynomial of five generators $V_1,V_2,V_3,W_1,Y_1$, where we assign
 the degree $1,2,3,1,1$ for $V_1,V_2,V_3,W_1,Y_1$ , respectively. 
The generators $V_1,V_2,V_3,W_1,Y_1$ are the functions of the moduli
parameter whose explicit forms are summarized in eqs.\eqref{gen-final}.

This fact provides a simple expression of the partition function of each genus;
This polynomial expression is completely closed and includes all the data of
the coefficients of instanton expansion. The polynomial is also more compact 
than the raw Feynman diagram expression; The number of terms grows only in 
power of the genus. 

The construction of this paper is as follows. In section \ref{sec:review},
we review the method of calculation of topological string amplitudes by
using the mirror symmetry and the holomorphic anomaly equation. In section 
\ref{sec:polynomial}, we prove that the partition function can be written as 
a polynomial of the five generators. Some of the coefficients are calculated in
section \ref{sec:coefficients}. Section \ref{sec:conclusion} is devoted to
conclusions and discussions. In appendix \ref{34}, polynomial form of
genus 3 and 4 partition function are written. In appendix \ref{app:2}, we
discuss the generalization to the Calabi-Yau hypersurfaces in weighted 
projective spaces treated in \cite{Klemm:1993tx}.

\section{Calculation of topological string amplitudes by mirror symmetry and
 the holomorphic anomaly equation}\label{sec:review}

In this section, we review the calculation of the topological string amplitudes
by the mirror symmetry\cite{Candelas:1991rm,Bershadsky:1993ta,Bershadsky:1994cx}. 
First, we explain the genus zero
 amplitudes following \cite{Candelas:1991rm}. 
After that, we will explain the genus one amplitudes and the higher genus ones
following \cite{Bershadsky:1993ta,Bershadsky:1994cx}. In this paper, we mainly
work with the quintic hypersurface in $\Cb P^4$. For this reason, we will 
concentrate to the case of quintic in the review in this section.

\subsection{Genus zero}

Let us review the genus zero amplitudes of the quintic\cite{Candelas:1991rm}.
The mirror manifold of the quintic is expressed by the orbifold of the 
hypersurface in $\Cb P^4$\cite{Greene:1990ud}
\begin{align}
 p:=x_1^5+x_2^5+x_3^5+x_4^5+x_5^5-5\psi x_1x_2x_3x_4x_5=0,
\end{align}
where $x_j,\ j=1,2,3,4,5$ are the homogeneous coordinates of $\Cb P^4$, and
$\psi$ is the moduli parameter. The orbifold group is $(\Zb_5)^3$ .
If we denote the
generators of this $(\Zb_5)^3$ by $g_1,g_2,g_3$, the action can be written as
\begin{align}
 g_j:x_j\to e^{\frac{2\pi i}{5}} x_j,\quad x_5\to e^{\frac{-2\pi i}{5}} x_5, \quad x_i\to x_i,\ (i\ne j,5).
\end{align}

We fix the gauge to the standard one in which the holomorphic 3-form $\Omega$ 
is written as 
\begin{align}
 \Omega=5\psi \frac{x_5 dx_1\wedge dx_2 \wedge dx_3}{\del p/\del x_4}.
\end{align}
In this gauge, the Picard-Fuchs equation for a period $w=\int \Omega$
is given by
\begin{align}
 \left\{\left(\psi \del_{\psi}\right)^4-\psi^{-5}
\left(\psi \del_{\psi}-1\right)
\left(\psi \del_{\psi}-2\right)
\left(\psi \del_{\psi}-3\right)
\left(\psi \del_{\psi}-4\right)
\right\}w=0. \label{PFeq}
\end{align}
There is a solution $\omega_0$ which is regular at $\psi \to \infty$. This 
solution is expressed by the expansion in $\psi^{-5}$ as
\begin{align}
 \omega_0(\psi)=\sum_{n=0}^{\infty}\frac{(5n)!}{(n!)^5 (5\psi)^{5n}}.\label{om0}
\end{align}
In order to write down the other solutions,
we extend the definition of $\omega_0$ to the function of $\psi$ and $\rho$ to
use the Frobenius argument. This function $\omega_0(\psi,\rho)$ should be the 
form 
\begin{align}
 \omega_{0}(\psi,\rho)=\sum_{n=0}^{\infty}\frac{\Gamma(5(n+\rho)+1)}{\Gamma(n+\rho+1)^5 (5\psi)^{5(n+\rho)}}.
\end{align}
The natural basis of the solutions are written as
\begin{align}
 \Pi=\left.\left(
\begin{array}{c}
 w_0\\
 w_1\\
 \del_{1}F_{0}\\
 \del_{0}F_{0}
\end{array}
\right)
=\left(\frac{2\pi i}{5}\right)^3
\left(
\begin{array}{c}
 \omega_0(\psi,\rho)\\
 \frac{1}{2\pi i} \del_{\rho}\omega_0(\psi,\rho)\\
 \frac52 \left(\frac{1}{2\pi i} \del_{\rho}\right)^2\omega_{0}(\psi,\rho)\\
 -\frac56 \left(\frac{1}{2\pi i} \del_{\rho}\right)^3\omega_{0}(\psi,\rho)
\end{array}
\right)\right|_{\rho=0}\; .
\end{align}
This basis are standard symplectic basis. Therefore,
the K\"ahler potential $K$ of the moduli space can be written as
\begin{align}
 e^{-K} = -i\Pi^{\dag} \Sigma \Pi,\qquad
\Sigma=\left(
\begin{array}{cccc}
 0 & 0 & 0 & 1 \\
 0 & 0 & 1 & 0 \\
 0 & -1& 0 & 0 \\
 -1& 0 & 0 & 0 \\
\end{array}
\right). \label{kpot}
\end{align}
This matrix $\Sigma$ is the ordinary symplectic bilinear form.
The metric of the moduli space is obtained as $G_{\psi\psib}=\del_{\psi} 
\delb_{\psib} K$.

Let us denote the complexified K\"ahler parameter in the A-model
 picture by $t$. The relation 
between $\psi$ and $t$ (``mirror map'') is given by
\begin{align}
 t=-2\pi i \frac{w_1}{w_0}
=-\log (5\psi)^{-5}-\frac{5}{\omega_0}\sum_{m=1}^{\infty}
\frac{(5m)!}{(m!)^5 (5\psi)^{5m}}(\Psi(1+5m)-\Psi(1+m)),\label{mm}
\end{align}
where $\Psi(x):=\del_{x}\log \Gamma(x)$ .

Another important observable is the Yukawa coupling. It is determined by
\begin{align}
 C_{\psi\psi\psi}=\Pi^{T} \Sigma \del_{\psi}^3 \Pi.
\end{align}
This equation and the Picard-Fuchs equation read the following differential 
equation for the Yukawa coupling.
\begin{align}
 \del_{\psi} C_{\psi\psi\psi}
  = \frac{2 \psi^{-1}+4 \psi^3}{1-\psi^5}C_{\psi\psi\psi}.
\end{align}
This differential equation can be solved as
\begin{align}
 C_{\psi\psi\psi}=\frac{(2\pi i)^3}{5^3}\frac{\psi^2}{1-\psi^5},
\label{yukawa}
\end{align}
where the normalization is fixed by the asymptotic behavior.
The Yukawa coupling in the $t$-frame becomes
\begin{align}
 C^{\spe}_{ttt}=\left(\frac{(2\pi i)^3}{5^7} \omega_{0}^2\right)^{-1}
\left(\frac{\del \psi}{\del t}\right)^3C_{\psi\psi\psi}.
\end{align}
The first factor in the right-hand side comes from the gauge transformation,
and the second factor is the contribution of the coordinate transformation. The 
$C^{\spe}_{ttt}$ gives the instanton expansion of the A-model picture and 
includes the information of the number of rational curves in the quintic.


These quantities, K\"ahler potential, metric, and Yukawa coupling are essential
to compute the higher genus amplitudes.

\subsection{Genus one and higher}

The one point function $\del_{\psi}F_1$ of genus one satisfies the holomorphic 
anomaly equation\cite{Bershadsky:1993ta}
\begin{align}
  \delb_{\psib}\del_{\psi}F_{1}=\frac12 C_{\psi\psi\psi}\bC_{\psib}^{\psi\psi}
   -\left(\frac{\chi}{24}-1\right)G_{\psi\psib},\qquad \chi=-200, \label{hae1}
\end{align}
where $\bC_{\psib}^{\psi\psi}$ is defined as
\begin{align}
 \bC_{\psib}^{\psi\psi}:=\bC_{\psib\psib\psib}(G_{\psi\psib})^{-2}e^{2K}.
\end{align}
Eq. \eqref{hae1} can be solved as
\begin{align}
 \del_{\psi} F_{1}= \frac12 \del_{\psi}\log\left[ (G_{\psi\psib})^{-1}
\exp\left(\frac{62}{3}K\right) \psi^{62/3}(1-\psi^5)^{-1/6}
\right].\label{g1res}
\end{align}
The holomorphic ambiguity is fixed by the asymptotic behavior.
In the $t$-frame, and topological limit ($\bar t\to \infty$), this one point 
function becomes
\begin{align}
 \del_{t}F^{\spe}_{1}&=\lim_{\bar t \to \infty}\frac{\del \psi}{\del t}
            \del_{\psi} F_{1}.
\end{align}
This function gives the instanton expansion in A-model picture, and includes the 
information of the number of elliptic curves in the quintic.

Let us turn to the $g\ge 2$ amplitudes. First, we introduce some notations.
We denote the vacuum bundle by $L$. Holomorphic 3-form $\Omega$ is a section
of $L$. For a section of $L$, the action of the gauge transformation (K\"ahler 
transformation) are parametrized by a holomorphic function $f(\psi)$ and 
expressed as
\begin{align}
 K(\psi,\psib)\to K(\psi,\psib)-\log f(\psi)-\log \bar f(\psib),\qquad
 \Omega\to f(\psi)\Omega.
\end{align}
The genus $g$ partition function $F_{g}$ is a section of $L^{2-2g}$ and 
transform as $F_g\to f(t)^{2-2g} F_g$.
Besides this symmetry of K\"ahler transformation, there is another gauge 
symmetry --- the reparametrization of the moduli. We will define the covariant 
derivative $D_{\psi}$ for these two gauge transformations.
If $h(\psi)$ is a section of $(T^{*})^m\otimes L^n$, the covariant derivative of 
$h$ is defined as  
\begin{align}
 D_{\psi}h=\del_{\psi}h+m\Gamma^{\psi}_{\psi\psi}h+n(\del_{\psi}K)h,
\label{cov-der1}
\end{align}
where $\Gamma^{\psi}_{\psi\psi}=-(G_{\psi\psi})^{-1}\del_{\psi} G_{\psi\psi}$ is
the Christoffel symbol.

Next, we consider the holomorphic anomaly equation.
The holomorphic anomaly equation for the 
genus $g$ partition function $F_g$ is given by \cite{Bershadsky:1994cx}
\begin{align}
 \delb_{\psib} F_{g}=\frac12 \bC_{\psib}^{\psi\psi}\left(D_{\psi}D_{\psi} 
 F_{g-1}+\sum_{r=1}^{g-1}D_{\psi}F_{\psi} D_{\psi} F_{g-r}\right). \label{hae}
\end{align}
A solution of \eqref{hae} is given by the Feynman rule as in 
\cite{Bershadsky:1994cx}. We denote this solution by $F_g^{(FD)}$.
The Feynman rule is composed of two kind of things: propagators and vertices.
We begin with the propagators.
It is useful to introduce the following quantities.
\begin{subequations}
 \label{s}
\begin{align}
& S^{\psi\psi}=\frac{1}{C_{\psi\psi\psi}}
        [2\del_{\psi}\log(e^{K}|f|^2)-\del_{\psi} \log(|v|^2 G_{\psi\psib})],\\
& S^{\psi}=\frac{1}{C_{\psi\psi\psi}}
      \left[\left(\del_{\psi}\log (e^{K}|f|^2)\right)^{2}
 -v^{-1}\del_{\psi}\left(v\del_{\psi} \log(e^{K}|f|^2)\right)
\right],\\
& S=\left[S^{\psi}-\frac12 D_{\psi}S^{\psi\psi}
                  -\frac12 (S^{\psi\psi})^2C_{\psi\psi\psi}
\right]\del_{\psi}\log(e^{K}|f|^2)+\frac12 D_{\psi} S^{1}
  +\frac12 S^{\psi\psi}S^{\psi}C_{\psi\psi\psi}, 
\end{align}
\end{subequations}
where $f$ is a holomorphic section of $L$ and $v$ is a holomorphic vector field 
on the moduli space.
In our gauge, we can set $f=\psi$ and $v=1$. The quantities $S^{\psi\psi}, 
S^{\psi}, S$ in \eqref{s} are determined to satisfy the relations
\begin{align}
 \bC_{\psib}^{\psi\psi}=\delb_{\psib}S^{\psi\psi},\qquad
 S^{\psi\psi}=(G_{\psi\psib})^{-1}\delb_{\psib}S^{\psi},\qquad
 S^{\psi}=(G_{\psi\psib})^{-1}\delb_{\psib}S.\label{sdef}
\end{align}
To show these relations, we use the special geometry relations\cite{Cremmer:1985hj}
\begin{equation}\label{sgeom}
\begin{split}
 \delb_{\psib}C_{\psi\psi\psi}=0,\qquad \del_{\psi}\bC_{\psi\psi\psi}=0,\\
 R_{\psi\psib\psi}{}^{\psi}:=-\delb_{\psib}\Gamma^{\psi}_{\psi\psi}
=2G_{\psi\psib}-C_{\psi\psi\psi}\bC_{\psib}^{\psi\psi}.
\end{split} 
\end{equation}
There are three types of propagators; The one connecting two solid 
lines, the one connecting solid and dashed lines, and the one connecting two 
dashed lines. The value of these propagators are written as
\begin{align}
 \includegraphics[width=2cm]{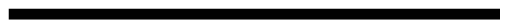} = -S^{\psi\psi},\qquad
 \includegraphics[width=2cm]{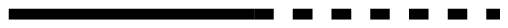}=-S^{\psi},\qquad
 \includegraphics[width=2cm]{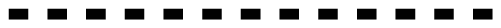} =-2S.
\end{align}

Let us turn to the vertices. A vertex
is labeled by three integers $g,n,m$.  $n$ solid lines and $m$ dashed 
lines end to the vertex labeled by $g,n,m$.
We denote the value of the vertex as
\begin{align}
\begin{minipage}{4cm}
  \includegraphics[width=3.5cm]{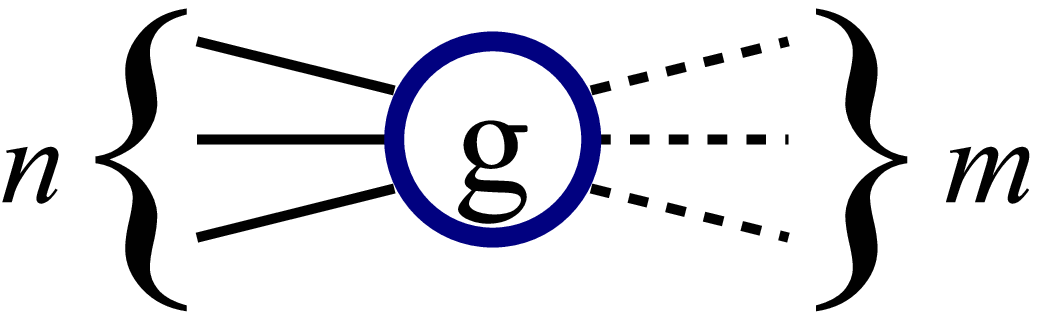}
\end{minipage}
= \lambda^{2g-2+n+m}\Ct^{(g)}_{\psi^n,\vp^m},
\end{align}
where $\lambda$ is the topological string coupling constant.
 The detail of the vertex
$\Ct^{(g)}_{\psi^n,\vp^m}$ are described as
\begin{equation}\label{defv}
\begin{split}
  \Ct^{(g)}_{\psi^n,\vp^{m+1}}
    =(2g-2+n+m)\Ct^{(g)}_{\psi^n,\vp^{m}},\qquad
 \Ct^{(g)}_{\psi^n,\vp^0}=C^{(g)}_{\psi^n},\ (2g-2+n \ge 1),\\
 C^{(g)}_{\psi^n}= D_{\psi}^n F_{g},\ (g \ge 2),\qquad
 C^{(1)}_{\psi^n}= D_{\psi}^{n-1} \del_{\psi} F_{1},\qquad
 C^{(0)}_{\psi^n}= D_{\psi}^{n-3} C_{\psi\psi\psi}\\
 \Ct^{(1)}_{\vp}=\frac{\chi}{24}-1,\quad \Ct^{(1)}_{\psi^{0}}=0,\qquad
 \Ct^{(0)}_{\psi\psi\vp^{m}}=\Ct^{(0)}_{\psi\vp^{m}}=\Ct^{(0)}_{\vp^{m}}=0.
\end{split}
\end{equation}

\begin{figure}
 \begin{center}
  \includegraphics[width=6cm]{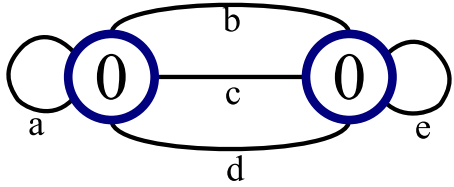}
 \end{center}
 \caption{A diagram which contribute to $F_{4}$ the symmetric factor
 is $s_{\cal D}=2\times 2 \times 3! \times 2=48$.}
 \label{fig1}
\end{figure}

The value of a diagram is obtained by multiplying the values of all the elements.
For example, the following diagram is the one which contribute to $F_5$. We can 
evaluate this diagram as
\begin{align}
\begin{minipage}{3.5cm}
  \includegraphics[width=3cm]{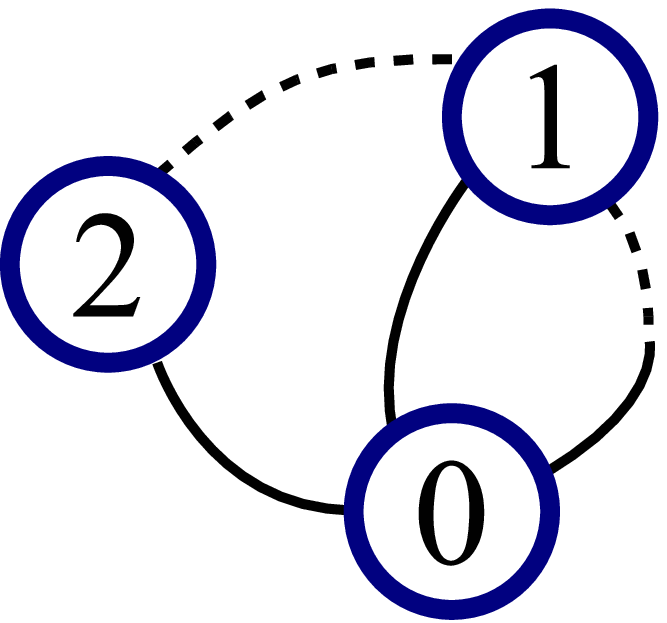}
\end{minipage}
 =\lambda^{8}\Ct^{(2)}_{\psi,\vp}
\Ct^{(0)}_{\psi\psi\psi}
\Ct^{(1)}_{\psi,\vp\vp}(-S^{\psi\psi})^2(-S^{\psi})(-2S).
\end{align}
The Feynman diagram part of the partition function is the sum of all connected 
diagrams $\Dcal$ divided by the appropriate symmetric factor (constant)
 $s_{\Dcal}$
\begin{align}
 F^{(FD)}_{g}\lambda^{2g-2}=- \sum_{\substack{\Dcal : \text{connected diagrams}\\ 
 \text{of order $\lambda^{2g-2}$}}} s_{\Dcal}^{-1}\; \Dcal.
\end{align}
Here, the symmetric factor $s_{\Dcal}$
 is the order of the symmetry group of the diagram ${\Dcal}$. For example,
figure \ref{fig1} is a diagram which contribute to $F_{4}$. The symmetric factor
is counted as follows; factor $2$ from the exchange
 of the two ends of the line $a$, factor $2$ from 
the exchange of the two ends of the line $e$, factor
$3!$ from the interchange of the three lines $b,c,d$, and factor $2$ from the 
left-right flip. The symmetric factor becomes $s_{\cal D}=2\times 2 \times 3! \times 2=48$.

Finally let us mention the holomorphic part.
The general solution of the holomorphic anomaly equation \eqref{hae} is
\begin{align}
 F_{g}(\psi,\psib)=F_{g}^{(FD)}(\psi,\psib)+f_{g}(\psi),
\end{align}
where $f_g(\psi)$ is a holomorphic function. From the asymptotic behavior,
$f_{g}(\psi)$ can be written as the following form.
\begin{align}
 f_{g}(\psi)=\sum_{j=1}^{[(2g-2)/5]}b_{g,j}\frac{1}{\psi^{5j}}
 +\sum_{j=0}^{2g-2}a_{g,j}\frac{1}{(1-\psi^5)^{j}}.
\label{holo-amb}
\end{align}
Here $[\cdot]$ denotes the Gauss symbol.
The coefficients $b_{g,j}$ are determined to cancel the singularity at $\psi=0$.
On the other hand, coefficients $a_{g,j}$ are ambiguities and should be 
determined by other information. 

The expression of a partition function in the $t$-frame is given by
\begin{align}
 F_{g}^{\spe}=\lim_{\bar t\to \infty}
     \left(\frac{(2\pi i)^3}{5^7} \omega_{0}^2\right)^{g-1} F_{g}.
\label{fgspe}
\end{align}
By expanding $F_{g}^{\spe}$ in terms of $e^{-t}$, we will obtain the instanton 
expansion of the A-model.
\section{Generators for the higher genus amplitudes}
\label{sec:polynomial}
In this section, we will show that the higher genus amplitudes are expressed
as polynomials of finite number of generators. First, in subsection \ref{sec:inf}
we introduce infinite number of generators, and show that the amplitudes can be 
written as polynomials of these generators. Second,
 in subsection \ref{sec:finite},
 these infinite number of generators turn out to be written as polynomials of finite number of generators.
Finally, in section \ref{sec:final}, we will reconsider the holomorphic anomaly
equation and show that the number of generators for partition functions reduces 
by one. We also state the final form of the claim in the $t$-frame.

\subsection{Expression of amplitudes by infinite number of
 generators}\label{sec:inf}
Let us introduce some notations.
\begin{align}
& A_p:=\frac{(\psi\del_{\psi})^p G_{\psi\psib}}{G_{\psi\psib}},\qquad
B_p:=\frac{(\psi\del_{\psi})^p e^{-K}}{e^{-K}},\qquad (p=1,2,3,\dots),\nn\\
&C:=C_{\psi\psi\psi}\psi^3,\qquad X:=\frac{1}{1-\psi^5}. \label{ABCX}
\end{align}
Especially, $A:=A_1$ and $B:=B_1$ are ``connections'' as
\begin{align}
 A:=A_1=-\psi \Gamma^{\psi}_{\psi\psi},\qquad B:=B_1=-\psi\del_{\psi}K.
\end{align}
We also denote amplitudes in ``(Yukawa coupling)$=1$ frame'' by
\begin{align}
 P_{g}:=C^{g-1}F_{g},\qquad
 P_{g}^{(n)}=C^{g-1} \psi^n C^{(g)}_{\psi^n},
\end{align}
where $P_g$ is defined for $g\ge 2$ and $P^{(n)}_{g}$ is defined
for $g=0$ and $n\ge 3$,  $g=1$ and $n\ge 1$, or $g\ge 2$ and $n\ge 0$.
The first thing we want to show is
\begin{proposition}\label{prop:inf}
  Each $P_g^{(n)}$ is an degree $(3g-3+n)$ inhomogeneous polynomial of
  $A_p,B_p,X,\ (p=1,2,3,\dots)$, where we assign
  ``degree'' $p$ to $A_p$ and $B_p$, and $1$ to $X$.
\end{proposition}
Now, we prove this statement. As preliminaries, we consider two things: the 
derivatives of generators and the expression of propagators.
The derivatives of the quantities of eq.\eqref{ABCX} becomes
\begin{align}
 \psi\del_{\psi}A_{p}=A_{p+1}-A A_{p},\quad 
 \psi\del_{\psi}A_{p}=A_{p+1}-A A_{p},\quad \psi\del_{\psi}X=5X(X-1),\quad
 \psi\del_{\psi}C=5XC.
\end{align}
We find two facts from these equations. First, if $h(A_p,B_p,X)$ is a polynomial
of $A_p,B_p,X$, then the derivative $\psi\del_{\psi} h$ is again a
polynomial of $A_p,B_p,X$. Second, the derivative $\psi\del_{\psi}$ increases 
the degree by $1$ in general. We can derive the similar facts for the covariant 
derivative $\psi D_{\psi}$. Let $h$ be a section of
 $(T^{*})^{\ell}\otimes L^{m}$, and assume $\psi^{\ell} C^{-m/2}h$ is a 
 polynomial of $A_p,B_p,X$ of degree $n$. Then the covariant derivative \eqref{cov-der1} of $h$
 becomes
\begin{align}
 C^{-m/2}\psi^{\ell+1}D_{\psi}h=\psi\del_{\psi} (\psi^{\ell} C^{-m/2}h)
 +[-\ell (A+1)-m (B-\frac{5}{2} X)](\psi^{\ell} C^{-m/2}h), \label{cov-der2}
\end{align}
and therefore $C^{-m/2}\psi^{\ell+1}D_{\psi}h$ is a polynomial
of $A_p,B_p,X$ of degree $(n+1)$ .

We can write the propagators
in eqs. \eqref{s} in terms of $A_p,B_p,X$ as
\begin{equation}\label{texp}
 \begin{split}
 T^{\psi\psi}:=&\frac{C}{\psi^2}S^{\psi\psi}=2-2B-A,\qquad
 T^{\psi}:=\frac{C}{\psi}S^{\psi}=2-3B+B_2,\\
 T:=CS=&\frac12\big[2 + 2A + A_2 - 3 B - A_2 B - B^2 - 
  2AB^2 + 2B^3 + 4B_2 - \\
  & \qquad 5BB_2 + B_3 - 5AX - 5BX + 
  5ABX + 10B^2X - 5B_2X\big].
\end{split}
\end{equation}
These equations explicitly shows
that $T^{\psi\psi},T^{\psi},T$ are inhomogeneous
polynomials of degree $1,2,3$ respectively.

Let us prove proposition \ref{prop:inf} by induction. If we assume $P_{g}^{(n)}$ 
is a polynomial of $A_p,B_p,X$ of degree $3g-3+n$, then $P_{g}^{(n+1)}$ can be 
written as
 \begin{align}
 P_{g}^{(n+1)}=\psi\del_{\psi} P_{g}^{(n)}
 +[-n(A+1)- (2-2g) (B-\frac{5}{2} X)]P_{g}^{(n)}, 
\end{align}
and $P_{g}^{(n+1)}$ turn out to be a polynomial of degree $(3g-3+n+1)$.
As for $g=0$, because $P^{(3)}_{g=0}=1$ by definition, we can conclude that
each $P^{(n)}_{g=0},\ n=3,4,5,\dots$ is a polynomial
of degree $(-3+n)$. In the case of $g=1$, eq. (\ref{g1res}) reads
\begin{align}
 P^{(1)}_{g=1}=\frac{31}{3}(1-B)
+\frac{5}{12}(X-1)-\frac12 A,
\end{align}
and we also find that each $P^{(n)}_{g=1},\ (n=1,2,3,\dots)$ is a polynomial of
degree $n$.

Let us fix $g\ge 2$ and assume each $P^{(n)}_{r},\ r<g$
is a polynomial of degree $(3r-3+n)$. In order to show $P_g$ to be a polynomial,
we pick up a diagram $\Dcal$ which contribute to $F_{g}$. We denote the number 
of vertices by $k$, the number of solid lines by $e$, the number of half-dashed 
lines by $e'$, and the number of dashed lines by $e''$ in the diagram $\Dcal$.
 We label each vertex by 
$j,\ (j=1,\dots,k)$ and let the genus of the vertex be $g_j$. We also let $n_j$ 
solid lines and $m_j$ dashed lines end on the $j$-th vertex. Then 
considering the number of lines, we find the relations
\begin{align}
 \sum_{j=1}^{k}n_{j}=2e+e',\qquad \sum_{j=1}^{k}m_j=e'+2e''.\label{rve}
\end{align}
Since $\Dcal$ contribute to $F_g$, we obtain the relation by counting the order 
of $\lambda$.
\begin{align}
 \sum_{j=1}^{k}(g_j-1)+e+e'+e''=g-1. \label{rg}
\end{align}
By using these relations and the expressions of vertices and propagators 
\eqref{defv}, \eqref{texp}, $\Dcal$ is evaluated as
\begin{align}
 \Dcal&=(\text{constant})\times \lambda^{2g-2}
\left(\prod_{j=1}^{k}\tilde{C}^{(g_j)}_{\psi^{n_j},\vp^{m_j}}\right)
(S^{\psi\psi})^{e}
(S^{\psi})^{e'}
(S)^{e''}\nn\\
&=(\text{constant})\times \lambda^{2g-2} \frac{1}{C^{g-1}}
\left(\prod_{j=1}^{k}P_{g_j}^{(n_j)}\right)
(T^{\psi\psi})^{e}
(T^{\psi})^{e'}
(T)^{e''}.
\end{align}
$T^{\psi\psi},T^{\psi},T$ are polynomials of $A_p,B_p,X$ because of 
eqs.\eqref{texp}, and $P_{r_j}^{(n_j)}$'s are also polynomials due to
the assumption of induction. Consequently, $C^{g-1}\Dcal$ is a polynomial.
Its degree is evaluated by using eqs.\eqref{rve},\eqref{rg} as
\footnote{Actually, we need a special care to the vertex with $g_j=1,n_j=0$.
The easiest way is to set $P^{(n=0)}_{g=1}=1$ temporally. The statement itself 
is correct.}
\begin{align}
 \sum_{j=1}^{k}(3g_j-3+n_j)+e+2e'+3e''=3g-3.
\end{align}
As a result, we can conclude that $C^{g-1}\Dcal$ is a polynomial of
$A_p,B_p,X$ of degree $(3g-3)$. 

So far, we have shown that the Feynman diagram
part of $P_{g}$ is a degree $(3g-3)$ polynomial. Now let us turn to the
holomorphic part. Eq. \eqref{holo-amb} is written by using $X$ as
\begin{align}
 f_{g}(\psi)=\sum_{j=1}^{[(2g-2)/5]}b_{g,j}\left(\frac{X}{X-1}\right)^j
 +\sum_{j=0}^{2g-2}a_{g,j} X^{j}.
\end{align}
Actually $C$ can be written as $C=\frac{(2\pi i)^3}{5^3}(X-1)$ 
because of the explicit form of the $C_{\psi\psi\psi}$ in \eqref{yukawa}.
Consequently, we can conclude that $C^{g-1}f_{g}$ is a degree $(3g-3)$ 
polynomial of $X$. Here, we have proved proposition \ref{prop:inf}. 

\subsection{Relation between generators}\label{sec:finite}

In this subsection, we will show that among the generators in \eqref{ABCX},
 $A_p\ (p=2,3,4,\dots)$ and $B_p,\ (p=4,5,6,\dots)$ are written as polynomials 
 of $A,B,B_2,B_3,X$. If we combine this fact and proposition \ref{prop:inf},
we can conclude that each $P^{(n)}_{g}$ is a degree $(3g-3+n)$ polynomial of $A,B,B_2,B_3,X$.

First, let us begin with $B_p$. By using eq.\eqref{kpot} and the definition
\eqref{ABCX}, we can write $B_p$ in the following form.
\begin{align}
 B_{p}=\frac{\Pi^{\dag}\Sigma (\psi\del_{\psi})^{p}\Pi}{\Pi^{\dag}\Sigma \Pi}.
\end{align}
Since each component of $\Pi$ is a period, $\Pi$ satisfies the Picard-Fuchs 
equation \eqref{PFeq}
\begin{align}
 \left\{\left(\psi \del_{\psi}\right)^4-\psi^{-5}
\left(\psi \del_{\psi}-1\right)
\left(\psi \del_{\psi}-2\right)
\left(\psi \del_{\psi}-3\right)
\left(\psi \del_{\psi}-4\right)
\right\}\Pi=0.
\end{align}
This equation reads the relation between generators
\begin{align}
 B_4=10 X B_3 -35 X B_2+50X B-24X.\label{br}
\end{align}
If we differentiate \eqref{br} and use the relation 
$\psi\del_{\psi}B_{p}=B_{p+1}-BB_{p}$ and \eqref{br} recursively,
we will obtain the expressions of $B_p,\ p=4,5,6,\dots$ in terms of
polynomials of $B,B_2,B_3,X$ of appropriate degrees.

Next, we turn to $A_p,\ (p=2,3,4,\dots)$. We can rewrite one of the special 
geometry relation $\del_{\psi}\bC_{\psib\psib\psib}=0$ by using the first 
equation of \eqref{sdef}, and the definition of $A,B$ \eqref{ABCX} as
\begin{align}
 2B\delb_{\psib}S^{\psi\psi}+2A \delb_{\psib}S^{\psi\psi}
 +\delb_{\psib}(\psi\del_{\psi}S^{\psi\psi})=0.
\end{align}
Moreover, multiply $C/\psi^2$ this equation and use eq.\eqref{texp},
then we obtain the following differential equation
\begin{align}
 -2A\delb_{\psib}B+\delb_{\psib}\left[
-2B^2-A^2-2AB+\frac{C}{\psi}\del_{\psi}S^{\psi\psi}\right]=0.\label{a20}
\end{align}
The last term inside the $\delb_{\psib}$ can be expressed in terms of 
$A,B,\dots$ 
\begin{align}
 \frac{C}{\psi}\del_{\psi}S^{\psi\psi}=-2B_2+2B^2-A_2+A^2-(5X-2)(2-2B-A).
\end{align}
We can also derive the following relation from the definition \eqref{ABCX}
\begin{align}
 -A\delb_{\psib}B=\delb(-B_2+B+B^2).\label{b20}
\end{align}
If we put these things into eq.\eqref{a20}, we obtain the differential equation
\begin{align}
  \delb_{\psib}(-4B_2-A_2-2AB-2B+2B^2-2A+10XB+5XA)=0.
\end{align}
We can fix the ``holomorphic ambiguity'' by asymptotic behavior, and obtain
the relation
\begin{align}
 A_2=-4B_2-2AB-2B+2B^2-2A+10XB+5XA-5X-1.\label{ar}
\end{align}
If we differentiate \eqref{ar}, and use \eqref{br} and \eqref{ar} recursively,
we will obtain the expression of $A_p,\ p=2,3,4,\dots$ as polynomials of
$A,B,B_2,B_3,X$ of appropriate degrees.

\subsection{Back to holomorphic anomaly equation}\label{sec:final}

Now, we have shown that each $P_{g}^{(n)}$ is a degree $(3g-3+n)$ polynomial of
$A,B,B_2,B_3,X$. In this subsection, we will rewrite the holomorphic anomaly 
equation \eqref{hae}, and see the nature of the polynomial $P_g(A,B,B_2,B_3,X)$.
As we will see, $P_g$ depends on some special combinations of $A,B,B_2,B_3,X$.

First, we multiply $C^{g-1}$ both side of eq.\eqref{hae} and see the 
left-hand side. Since $X$ is holomorphic, the anti-holomorphic derivative of 
$P_g$ becomes
\begin{align}
 \delb_{\psib}P_g=
\delb_{\psib}A \deldel{P_g}{A}
+\delb_{\psib}B \deldel{P_g}{B}
+\delb_{\psib}B_2 \deldel{P_g}{B_2}
+\delb_{\psib}B_3 \deldel{P_g}{B_3}.
\end{align}
As we have seen in eq.\eqref{b20}, $\delb_{\psib}B_2$ can be written as
\begin{align}
 \delb_{\psib}B_2=(A+1+2B)\delb_{\psib}B.
\end{align}
Similarly, we can rewrite $\delb_{\psib}B_3$ as
\begin{align}
\delb_{\psib}B_3=\{(B+5X)(1+A+2B)-B_2-10X\}\delb_{\psib}B.
\end{align}

If we put these things into eq.\eqref{hae} and use
 the first equation of \eqref{sdef} and eq.\eqref{texp},
the holomorphic anomaly equation can be written as
\begin{align}
 \delb_{\psib}A \frac{\del P_g}{\del A}
 +\delb_{\psib}B \left[
\frac{\del P_g}{\del B}
+(A+1+2B)\frac{\del P_g}{\del B_2}
+\{(B+5X)(1+A+2B)-B_2-10X\}\frac{\del P_g}{\del B_3}
\right]\nn\\
=\frac12(-\delb_{\psib} A-2\delb_{\psib}B)\left(
P_{g-1}^{(2)}+\sum_{r=1}^{g-1}P_{r}^{(1)}P_{g-r}^{(1)}
\right)\label{hae26}
\end{align}

If we assume $\delb_{\psib}A$ and $\delb_{\psib}B$ are independent, 
eq.\eqref{hae26} yields
two independent differential equations. One of these is written as
\begin{align}
\left[-2\frac{\del}{\del A}+
\frac{\del}{\del B}
+(A+1+2B)\frac{\del}{\del B_2}
+\{(B+5X)(1+A+2B)-B_2-10X\}\frac{\del}{\del B_3}
\right]P_g=0.\label{pg}
\end{align}
This differential equation gives a constraint for the partition function $P_g$.
To see this, it is convenient to change the variables from $(A,B,B_2,B_3,X)$
to $(u,v_1,v_2,v_3,X)$ as 
\begin{equation}
 \begin{split}
 &u=B,\qquad
v_1=A+1+2B,\qquad
 v_2=B_2-B(A+1+2B),\\
 &v_3=B_3-B\{B(1+A+2B)-B_2+5X(1+A+2B)-10X\},
\end{split}
\end{equation}
or
\begin{equation}
 \begin{split}\label{uvsystem}
 &B=u,\qquad A=v_1-1-2u,\qquad
 B_2=v_2+uv_1,\\& B_3=v_3+u(-v_2+5X(v_1-2))
\end{split}
\end{equation}
In variables $(u,v_1,v_2,v_3,X)$, eq.\eqref{pg} simplifies to
\begin{align}
 \frac{\del P_g}{\del u}=0.
\end{align}
As a result, we can conclude that $P_g$ is independent of $u$ in the valuable 
$(u,v_1,v_2,v_3,X)$. We summarize this result as the following proposition.
\begin{proposition}\label{prop:semi-final}
 Each $P_g,\ g=2,3,4,\dots$ is a degree $(3g-3)$ inhomogeneous 
polynomial of $v_1,v_2,v_3,X$, where we assign
 the degree $1,2,3,1$ for $v_1,v_2,v_3,X$, respectively.
\end{proposition}

Finally, we state proposition \ref{prop:semi-final} in the A-model picture.
Recall that the partition function $F_g^{\spe}$ in A-model picture is related
to $P_g$ by
\begin{align}
 F_g^{\spe}=\lim_{\bar t\to \infty}
   \left(\frac{(2\pi i)^3\omega_{0}^2}{5^7 C}\right)^{g-1}P_g.
\end{align}
Therefore, if we define
\begin{align}
 W_{1}:=\left(\frac{(2\pi i)^3\omega_{0}^2}{5^7 C}\right)^{1/3},\qquad
 V_{j}:=\lim_{\bar t \to \infty} v_j W_1^{j},\ (j=1,2,3),\qquad
 Y_{1}:=XW_1,
\end{align}
then the final form of the claim is obtained as the theorem.
\begin{theorem}\label{thm:final}
 Each $F_g^{\spe},\ g=2,3,\dots$ is a degree $(3g-3)$ quasi-homogeneous 
polynomial of $V_1,V_2,V_3,W_1,Y_1$, where we assign
 the degree $1,2,3,1,1$ for $V_1,V_2,V_3,W_1,Y_1$ , respectively.
\end{theorem}

We write the summary of the final form of generators $V_1,V_2,V_3,W_1,Y_1$
here. We use the fact that  $\psib\to \infty $, $G_{\psi\psib}\propto \psib^{-2} 
\del_{\psi} t$ and $e^{-K}\to \omega_{0}$ in the limit $\bar t \to \infty$.
The function $\omega_0(\psi)$ and $t(\psi)$ is as written in eq.\eqref{om0} and
eq.\eqref{mm} respectively. The generators are expressed as
\begin{equation}\label{gen-final}
\begin{split}
 W_1=\left(\frac{\omega_0^2(\psi^{-5}-1)}{5^4}\right)^{1/3},\qquad
Y_1=W_1 \frac{1}{1-\psi^5},\\
 V_1=W_1\left(
 \frac{(\psi\del_{\psi})^2 t}{\psi\del_{\psi}t}+2\frac{\psi\del_{\psi}\omega_0}{\omega_0}\right),\qquad
V_2=W_1^2 \frac{(\psi\del_{\psi})^2 \omega_0}{\omega_0}-W_1V_1\frac{\psi\del_{\psi} \omega_0}{\omega_0},\\
V_3=W_1^3\frac{(\psi\del_{\psi})^3\omega_0}{\omega_0}-W_1\frac{\psi\del_{\psi}\omega_0}{\omega_0}(-V_2+5Y_1V_1-10W_1Y_1).
\end{split} 
\end{equation}

To obtain the instanton expansion, we need to write down the inverse relation
$\psi=\psi(t)$ as a power series of $e^{-t}$ and insert it to the above expressions.

\section{Some results for the coefficients of the polynomial representation}
\label{sec:coefficients}
So far, we have proved that $P_g$ is a polynomial of $v_1,v_2,v_3,X$. In this
section, we try to determine the coefficients of the polynomial. To do this, the 
most serious problem is the holomorphic ambiguity. As for some lower genus, say
genus 2,3, and 4, we can fix the ambiguity by known results 
\cite{Bershadsky:1994cx,Katz:1999xq}. 
There are also a part of the coefficients which do not suffer
 from the ambiguity. 
We will calculate some of these coefficients for all genus.

In this section, we use proposition \ref{prop:semi-final} form of $P_g$. In 
order to get theorem \ref{thm:final} form of $F_g^{\spe}$, replace $v_j$ with 
$V_j$ and $X$ with $Y_1$, and adjust the degree with $W_1$.
\subsection{Lower genus partition functions}
We can calculate the coefficients of the polynomial by holomorphic anomaly 
equation or equivalently the Feynman rule. We should fix the holomorphic 
ambiguity at each order. For example, the genus $2$ partition function can
be written in the polynomial form
\begin{align}
 P_2=&\frac{3125}{144} - \frac{15625}{288}v_1 + 
  \frac{125}{24}v_1^2 - \frac{5}{24}v_1^3 - 
  \frac{3125}{36}v_2 + 
  \frac{25}{6}v_1v_2 + 
  \frac{350}{9}v_3 - \frac{28795}{144}X - 
  \frac{835}{144} v_1 X \nn\\&+ \frac{5}{6}v^2 X - 
  \frac{2375}{12}v_2 X + 
  \frac{205}{144}X^2 - \frac{325}{288}v_1 X^2 + 
  \frac{25}{48}X^3.
\end{align}
We can also write the genus 3 and 4 partition function in the polynomial form,
and show them in appendix \ref{34}. 

\subsection{Coefficients of $v_3^n$}

We can calculate some simple part of the coefficients in the full order. In this 
subsection, we consider the coefficients of $v_3^n$ term. First, we define 
the following partition function
\begin{align}
 Z(\lambda,v_1,v_2,v_3,X)=\exp\left(\sum_{g=2}^{\infty}
   \lambda^{2g-2} P_g(v_1,v_2,v_3,X)\right).
\end{align}
The holomorphic anomaly equation can be written in the simple form as explained
in \cite{Bershadsky:1994cx}
\begin{align}
 \delb_{\psib}Z = \frac12 \lambda^{2}
(-\delb_{\psib}A-2\delb_{\psib}B ) \left[(P^{(2)}_{1}+(P^{(1)}_1)^2)Z
+2P^{(1)}_1\psi D_{\psi}Z+\psi^2D_{\psi}^2Z\right],\\
\psi D_{\psi}Z:=\psi \del_{\psi}Z+\left(u-\frac52 X\right)\lambda\del_{\lambda}Z.
\end{align}
The both side of this equation become quadratic in $u$.
If we use explicit form of $P_{1}^{(1)}$ and $P_{1}^{(2)}$,
and compare each coefficients of
$u$, we obtain three partial differential equation of Z
\begin{align}
 -\frac{2}{\lambda^2}\deldel{Z}{v_1}=&
 (\psi\del_{\psi})^2 Z + \frac{25}{4}X^2 (\lambda\del_{\lambda})^2Z-5X 
 \psi\del_{\psi}\lambda\del_{\lambda}Z 
 +\left(-2v_1+\frac{5}{6}X-\frac{25}{2}\right)\psi\del_{\psi}Z \nn\\&
 +\left(v_2+5v_1 X-\frac{175}{12}X^2+\frac{175}{4}X\right)\lambda\del_{\lambda}Z 
 \nn\\ &
 +\left(\frac{15625}{144}-\frac{125}{6}v_1 + \frac{5}{4}v_1^2-\frac{25}{3}v_2
 +\frac{835}{72}X-\frac{10}{3}v_1X+\frac{325}{144}X^2
\right)Z,
\end{align}
\begin{align}
 \frac{2}{\lambda^2}\left(\deldel{Z}{v_2}+5X\deldel{Z}{v_3}\right)
=&-\frac{50}{3}\psi\del_{\psi}Z+\left(\frac{85}{2}X-v_1-\frac{25}{2}\right)
   \lambda\del_{\lambda}Z+2\psi\del_{\psi}\lambda\del_{\lambda}Z
   -5X(\lambda\del_{\lambda})^2 Z \nn\\&
  +\left(-\frac{3125}{18}+\frac{25}{3}v_1-\frac{125}{18}X\right)Z,
\end{align}
\begin{align}
 \frac{2}{\lambda^2}\deldel{Z}{v_3}=&
 \left(\frac{\chi}{12}-1\right)\lambda\del_{\lambda}Z+(\lambda\del_{\lambda})^2 
 Z+\frac{\chi}{24}\left(\frac{\chi}{24}-1\right)Z.\label{pde}
\end{align}
Here $\psi\del_{\psi}$ act to Z as
\begin{align}
 \psi\del_{\psi}Z=&(-v_1^2-2 
 v_2-10X+5v_1X)\deldel{Z}{v_1}+(-v_1v_2+v_3)\deldel{Z}{v_2}\nn\\&
+(v_2^2-24X-25v_2X
 -5v_1v_2 X+10 v_3X)\deldel{Z}{v_3}+5X(X-1)\deldel{Z}{X}
\end{align}

Now, in order to see only the $v_3$ and $\lambda$ dependence, we define a
function
\begin{align}
 \Zt(\lambda,v_3):=Z(\lambda,v_1=0,v_2=0,v_3,X=0).
\end{align}
The coefficients of $v_3^n$ terms are encoded in this function $\Zt$.
This function $\Zt$ satisfies also the differential equation \eqref{pde}.
As a result, $\Zt$ is solved by the formal power series as
\begin{align}
 \Zt=\sum_{n=0}^{\infty}\sum_{k=0}^{n}\lambda^{2n}v_3^{k}
\frac{\Gamma\left(\frac{\chi}{24}-1+2n\right)}{2^k 
k!\Gamma\left(\frac{\chi}{24}-1+2n-2k\right)} \alpha_{n-k},
\label{v3res}
\end{align}
where the constants $\alpha_{\ell},\ (\ell=0,1,2,\dots)$ are part of the 
holomorphic ambiguities. These are fixed by considering the constant map 
contribution\cite{Marino:1998pg,Gopakumar:1998ii,Faber:1998}, namely
\begin{align}
 \lim_{t\to \infty} F_{g}^{\spe}=\frac{(-1)^g B_{g}B_{g-1}}{4g(2g-2)(2g-2)!}\chi,
\end{align}
where $B_g,\ g=1,2,3,\dots$ are the Bernoulli numbers. We also use the fact
that in the limit
$\bar t\to \infty$ and $t \to \infty$, $v_j$ and $X$ vanish. The 
$\alpha_{n}$ are expressed as 
\begin{align}
 \Zt(\lambda,v_3=0)=\sum_{n=0}^{\infty}\lambda^{2n}\alpha_{n}
=\exp\left(\sum_{g=2}^{\infty}\lambda^{2g-2}
\frac{(-1)^g B_{g}B_{g-1}}{ \dot 4g(2g-2)(2g-2)!}(-5^4\chi)
\right). 
\end{align}

Let us make a remark here. We denote the generating function of the coefficient 
of $\lambda^{2n}v_3^n$ in \eqref{v3res} by $\Zt^{(0)}$. The explicit form of 
$\Zt^{(0)}$ can be written as a formal series 
\begin{align}
  \Zt^{(0)}(\lambda^2 v_3)=\sum_{n=0}^{\infty}(\lambda^{2}v_3)^{n}
\frac{\Gamma\left(\frac{\chi}{24}-1+2n\right)}{2^n 
n!\Gamma\left(\frac{\chi}{24}-1\right)}.
\end{align}
This series can be rewritten as the asymptotic expansion of
Kummer confluent hypergeometric function ${}_1F_1(\alpha,\gamma;z)$.
We can write
\begin{align}
 \Zt^{(0)}(\lambda^2 v_3)
=C_{1} (2\lambda^2 v_3)^{-\frac12\left(\frac{\chi}{24}-1\right)}
{}_1F_1\left(\frac12\left(\frac{\chi}{24}-1\right),\frac12
   ;-\frac{1}{2\lambda^2v_3}\right)
+C_{2} (2\lambda^2 v_3)^{-\frac{\chi}{48}}
{}_1F_1\left(\frac{\chi}{48},\frac32
   ;-\frac{1}{2\lambda^2v_3}\right),\label{resum}
\end{align}
where $C_1$ and $C_2$ are constants which satisfies
\begin{align}
 \frac{\Gamma\left(\frac12\right)}{\Gamma\left(1 -   
\frac{\chi}{48}\right)} C_1
+ \frac{\Gamma\left(\frac32\right)}{\Gamma\left(\frac32 -
 \frac{\chi}{48}\right)} C_2=1.
\end{align}
The expression \eqref{resum} might give some
non-perturbative information of the topological string theory.
\section{Conclusion and Discussion}
\label{sec:conclusion}

In this paper, we have shown that the topological partition functions of the 
quintic can be written as polynomials of five generators. We have written down
the polynomial forms of $F_2,F_3,F_4$. We also obtain the coefficients of 
$v_3^n$ for all genus.

To fix the holomorphic ambiguity is the most serious problem to obtain the 
coefficients of the polynomial. One possible way to do this is using the 
heterotic dual 
description\cite{Antoniadis:1994ze,Antoniadis:1995zn,Serone:1997bk,Marino:1998pg,
Hosono:1999qc}. Also the large N duality \cite{Diaconescu:2003dk} might give some hints.

The fact that $F_g$'s are polynomials of five generators implies that there are
polynomial relations between $F_g$'s. In other words, for $2\le g_1 
<g_2<\dots<g_k$, $k\ge 6$, there is a quasi-homogeneous polynomial 
$Q(F_{g_1},\dots,F_{g_k})$ such that
\begin{align}
 Q(F_{g_1},\dots,F_{g_k})=0.
\end{align}
These polynomial relations are completely gauge invariant. Therefore we can 
expect some physical or mathematical meaning of the coefficients of this 
polynomial. If this meaning becomes clear, it might be useful to fix the 
holomorphic ambiguity.

In this paper, we mainly treat the quintic hypersurface. We can also do
the similar analysis for the Calabi-Yau hypersurfaces
in weighted projective spaces
treated in \cite{Klemm:1993tx}. See appendix \ref{app:2}. The generalization
to other Calabi-Yau manifolds, especially complete intersection in products
of weighted projective spaces \cite{Hosono:1995qy,Hosono:1995ax}
is a future problem.

\subsection*{Acknowledgment}
We would like to thank Jun Li, Bong H. Lian, Kefeng Liu, Hirosi Ooguri, and 
Cumrun Vafa for useful discussions.
This work was supported in part by NSF grants DMS-0074329 and DMS-0306600.

\appendix
\section{Polynomial form of genus 3 and 4 partition function}
\label{34}

Here we show the genus 3 and 4 partition functions in the polynomial form. We 
use the result of \cite{Katz:1999xq} to fix the ambiguity.

\begin{equation} 
\begin{split}
 P_3=&\frac{5}{72576}( 781250 - 2734375\,v_1^3 + 
      787500\,v_1^4 - 94500\,v_1^5 + 
      4536\,v_1^6 + 6562500\,v_2 - 
      16721250\,v_1\,v_2 \\&- 
      2625000\,v_1^2\,v_2 + 
      819000\,v_1^3\,v_2 - 
      54432\,v_1^4\,v_2 - 
      18112500\,v_2^2 - 
      1772400\,v_1\,v_2^2 \\&+ 
      295344\,v_1^2\,v_2^2 - 
      936320\,v_2^3 - 4935000\,v_3 + 
      12337500\,v_1\,v_3 - 
      1184400\,v_1^2\,v_3 + 
      47376\,v_1^3\,v_3 \\&+ 
      19740000\,v_2\,v_3 - 
      947520\,v_1\,v_2\,v_3 - 
      4421760\,v_3^2 + 27683000\,X  - 
      72635850\,v_1\,X \\&+ 
      12252135\,v_1^2\,X - 
      3366615\,v_1^3\,X + 
      604044\,v_1^4\,X - 
      41580\,v_1^5\,X - 
      81544680\,v_2\,X \\&- 
      54284034\,v_1\,v_2\,X + 
      3202584\,v_1^2\,v_2\,X + 
      93240\,v_1^3\,v_2\,X - 
      99165864\,v_2^2\,X + 
      3824016\,v_1\,v_2^2\,X \\&+ 
      45473064\,v_3\,X + 
      1318632\,v_1\,v_3\,X - 
      189504\,v_1^2\,v_3\,X + 
      45007200\,v_2\,v_3\,X \\&- 
      112828006\,X^2 - 12527550\,v_1\,X^2 + 
      5722185\,v_1^2\,X^2 - 
      1658685\,v_1^3\,X^2 + 
      176400\,v_1^4\,X^2 \\&- 
      233375520\,v_2\,X^2 - 
      3865134\,v_1\,v_2\,X^2 + 
      104160\,v_1^2\,v_2\,X^2 - 
      113818740\,v_2^2\,X^2 \\&- 
      323736\,v_3\,X^2 + 
      256620\,v_1\,v_3\,X^2 + 
      3339968\,X^3 - 4795350\,v_1\,X^3 + 
      2353785\,v_1^2\,X^3 \\&- 
      444325\,v_1^3\,X^3 + 
      819840\,v_2\,X^3 - 
      266910\,v_1\,v_2\,X^3 - 
      118440\,v_3\,X^3 + 1696500\,X^4 \\&- 
      1683150\,v_1\,X^4 + 
      686175\,v_1^2\,X^4 + 
      119700\,v_2\,X^4 + 477000\,X^5 - 
      598500\,v_1\,X^5 + 225000\,X^6 ) .
\end{split}
\end{equation}
{\tiny
\jot 0.5ex
\begin{equation} 
\begin{split}
 P_4=&\frac{1}{850500000000000}
   (476837158203125000 - 
    1251697540283203125\,v_1^5 + 
    640869140625000000\,v_1^6 - 
    144195556640625000\,v_1^7 \\&+ 
    16611328125000000\,v_1^8 - 
    815800781250000\,v_1^9 + 
    1907348632812500000\,v_2 - 
    91552734375000000\,v_1\,v_2 \\&+ 
    2002716064453125000\,v_1^2\,v_2 - 
    12067031860351562500\,v_1^3\,v_2 + 
    1625244140625000000\,v_1^4\,v_2 + 
    307617187500000000\,v_1^5\,v_2 \\&- 
    88593750000000000\,v_1^6\,v_2 + 
    6157265625000000\,v_1^7\,v_2 + 
    10013580322265625000\,v_2^2 - 
    23752212524414062500\,v_1\,v_2^2 \\&- 
    15252685546875000000\,v_1^2\, v_2^2 + 
    2514770507812500000\,v_1^3\, v_2^2 + 
    55371093750000000\,v_1^4\,v_2^2 - 
    20981953125000000\,v_1^5\,v_2^2 \\&- 
    22117333984375000000\,v_2^3 - 
    9664306640625000000\,v_1\,v_2^3 + 
    970429687500000000\,v_1^2\, v_2^3 \\&+ 
    27240937500000000\,v_1^3\,v_2^3 - 
    3177453125000000000\,v_2^4 + 
    38403750000000000\,v_1\,v_2^4 - 
    450134277343750000\,v_3 \\&- 
    4005432128906250000\,v_1\,v_3 + 
    5391311645507812500\,v_1^2\,v_3 + 
    3402587890625000000\,v_1^3\,v_3 - 
    1176635742187500000\,v_1^4\,v_3 \\&+ 
    147656250000000000\,v_1^5\,v_3 - 
    7190859375000000\,v_1^6\,v_3 - 
    16918945312500000000\,v_2\,v_3 + 
    43109472656250000000\,v_1\,v_2\,v_3 \\&+ 
   1896972656250000000\,v_1^2\,v_2\,v_3 - 
    1176328125000000000\,v_1^3\,v_2\,v_3 + 
    84223125000000000\,v_1^4\,v_2\,v_3 + 
    41825683593750000000\,v_2^2\,v_3 \\&+ 
    1608140625000000000\,v_1\,v_2^2\,v_3 -
     443480625000000000\,v_1^2\, v_2^2\,v_3 + 
    1499575000000000000\,v_2^3\,v_3 + 
    5387402343750000000\,v_3^2 \\&- 
    13468505859375000000\,v_1\,v_3^2 + 
    1292976562500000000\,v_1^2\,v_3^2 - 
    51719062500000000\,v_1^3\,v_3^2 - 
    21549609375000000000\,v_2\,v_3^2 \\&+ 
    1034381250000000000\,v_1\,v_2\,v_3^2 + 
    3218075000000000000\,v_3^3 + 
    3776550292968750000\,X + 
    17503967285156250000\,v_1\,X \\&- 
    22250582885742187500\,v_1^2\,X - 
    28078023681640625000\,v_1^3\,X + 
    14768562445068359375\,v_1^4\,X - 
    5377867954101562500\,v_1^5\,X \\&+ 
    1345452978515625000\,v_1^6\,X - 
    190480253906250000\,v_1^7\,X + 
    11572558593750000\,v_1^8\,X + 
    95444091796875000000\,v_2\,X \\&- 
    226548006591796875000\,v_1\,v_2\,X - 
    25295876586914062500\,v_1^2\,v_2\,X -
     14333584726562500000\,v_1^3\, v_2\,X +
    3472856542968750000\, v_1^4\,v_2\,X \\&+ 
    3725859375000000\,v_1^5\,v_2\,X - 
    32336718750000000\,v_1^6\,v_2\,X - 
    143172919189453125000\,v_2^2\,X - 
    216676193261718750000\,v_1\,v_2^2\,X \\&-
     13990677832031250000\,v_1^2\,v_2^2\,X + 
    5559784453125000000\,v_1^3\,v_2^2\,X - 
    248480859375000000\,v_1^4\,v_2^2\,X -
    217442333828125000000\,v_2^3\,X \\&- 
    10364004218750000000\,v_1\,v_2^3\,X +
    2042381250000000000\,v_1^2\, v_2^3\,X - 
    7434783125000000000\,v_2^4\,X - 
    59099116210937500000\,v_3\,X \\&+ 
    152810406738281250000\,v_1\,v_3\,X - 
    22033919824218750000\,v_1^2\,v_3\, X +
     5277016992187500000\,v_1^3\, v_3\,X -
    948538828125000000\,  v_1^4\,v_3\,X \\&+ 
    66002343750000000\,v_1^5\,v_3\,X + 
    175039331250000000000\,v_2\,v_3\,X + 
    122346708750000000000\,v_1\,v_2\,v_3\,X -
    8964263437500000000\,  v_1^2\,v_2\,v_3\,X \\&+ 
    2756250000000000\,v_1^3\,v_2\, v_3\,X  +
    217267594687500000000\, v_2^2\,v_3\,X - 
    8929869375000000000\,v_1\,v_2^2\, v_3\,X -
    49641680156250000000\,  v_3^2\,X \\&- 
    1439513906250000000\,v_1\,v_3^2\, X +
    206876250000000000\,v_1^2\, v_3^2\,X - 
    49133109375000000000\,v_2\,v_3^2\,X +
    163579760009765625000\,X^2 \\&- 
    443313000805664062500\,v_1\,X^2 + 
    101822160087890625000\,v_1^2\,X^2 - 
    48842653500976562500\,v_1^3\,X^2 + 
    19449174848632812500\,v_1^4\,X^2 \\&- 
    5720941560058593750\,v_1^5\,X^2 + 
    1020827636718750000\,v_1^6\,X^2 - 
    78894580078125000\,v_1^7\,X^2 - 
    304551039785156250000\,v_2\,X^2 \\&- 
    734736760839843750000\,v_1\,v_2\, X^2 +
    89751814687500000000\,v_1^2\, v_2\,X^2 - 
    18545016943359375000\,v_1^3\,v_2\, X^2 +
    1985324414062500000\,v_1^4\,  v_2\,X^2 \\&+ 
    25692187500000000\,v_1^5\,v_2\, X^2 -
    1024105277343750000000\,v_2^2\,X^2 - 
    276891787060546875000\,v_1\,v_2^2\,  X^2 +
    13071216210937500000\,v_1^2\,  v_2^2\,X^2 \\&+ 
    642981445312500000\,v_1^3\,v_2^2\, X^2 -
    545788993203125000000\,v_2^3\,X^2 + 
    19217201171875000000\,v_1\,v_2^3\, X^2 +
    248822582949218750000\,v_3\,X^2 \\&+ 
    23662275000000000000\,v_1\,v_3\,X^2 - 
    9440637158203125000\,v_1^2\,v_3\,  X^2 +
    2619044531250000000\,v_1^3\, v_3\,X^2 - 
    280534570312500000\,v_1^4\,v_3\, X^2 \\&+
    507984192187500000000\,v_2\, v_3\,X^2 + 
    10447962656250000000\,v_1\,v_2\, v_3\,X^2 - 
    743908593750000000\,v_1^2\,v_2\, v_3\,X^2 + 
    248917708593750000000\,v_2^2\,v_3\,X^2 \\&+
    353413593750000000\,v_3^2\,X^2 - 
    280144921875000000\,v_1\,v_3^2\, X^2 -
    396523506113281250000\,X^3 - 
    103635926660156250000\,v_1\,X^3 \\&+ 
    73077346025390625000\,v_1^2\,X^3 - 
    39332128759765625000\,v_1^3\,X^3 + 
    14423263732910156250\,v_1^4\,X^3 - 
    3344353051757812500\,v_1^5\,X^3 \\&+ 
    342342041015625000\,v_1^6\,X^3 - 
    1281315666738281250000\,v_2\,X^3 - 
    94647117539062500000\,v_1\,v_2\,X^3 + 
    33246505810546875000\,v_1^2\,v_2\, X^3 \\&-
    7335175585937500000\,v_1^3\, v_2\,X^3 + 
    262899902343750000\,v_1^4\,v_2\, X^3 -
    1293917594238281250000\,v_2^2\,X^3 - 
    16293676464843750000\,v_1\,v_2^2\, X^3 \\&-
    129414550781250000\,v_1^2\,v_2^2\,X^3 - 
    418847288671875000000\,v_2^3\,X^3 - 
    6279172187500000000\,v_3\,X^3 + 
    8323518164062500000\,v_1\,v_3\,X^3 \\&- 
    3734208105468750000\,v_1^2\,v_3\,X^3 +
    708045898437500000\,v_1^3\,v_3\,X^3 - 
    2425368750000000000\,v_2\,v_3\,X^3 + 
    1194670312500000000\,v_1\,v_2\,v_3\,X^3 \\&+ 
    129297656250000000\,v_3^2\,X^3 + 
    29053259482421875000\,X^4 - 
    55182304052734375000\,v_1\,X^4 + 
    46439463281250000000\,v_1^2\,X^4 \\&- 
    23409994384765625000\,v_1^3\,X^4 + 
    7303842895507812500\,v_1^4\,X^4 - 
    1051710479736328125\,v_1^5\,X^4 + 
    24480730859375000000\,v_2\,X^4 \\&- 
    27793218750000000000\,v_1\,v_2\,X^4 + 
    11772791748046875000\,v_1^2\,v_2\, X^4 -
    1098571899414062500\,v_1^3\, v_2\,X^4 + 
    2913170654296875000\,v_2^2\,X^4 \\&- 
    463520141601562500\,v_1\,v_2^2\, X^4 - 
    3067916894531250000\,v_3\,X^4 + 
    2679930175781250000\,v_1\,v_3\,X^4 - 
    1095309448242187500\,v_1^2\,v_3\, X^4 \\&-
    542390625000000000\,v_2\,v_3\, X^4 +
    17889105640561810592\,X^5 - 
    29737974121093750000\,v_1\,X^5 + 
    24661093139648437500\,v_1^2\,X^5 \\&- 
    10760107421875000000\,v_1^3\,X^5 + 
    2387512054443359375\,v_1^4\,X^5 + 
    9409335937500000000\,v_2\,X^5 - 
    9114737548828125000\,v_1\,v_2\,X^5 \\&+ 
    2014674682617187500\,v_1^2\,v_2\, X^5 +
    182186279296875000\,v_2^2\,X^5 - 
    761422851562500000\,v_3\,X^5 + 
    956535644531250000\,v_1\,v_3\,X^5 \\&+ 
    7895910935673253816\,X^6 - 
    15704566040039062500\,v_1\,X^6 + 
    10229498291015625000\,v_1^2\,X^6 - 
    3995840454101562500\,v_1^3\,X^6 \\&+ 
    2802546386718750000\,v_2\,X^6 - 
    1848229980468750000\,v_1\,v_2\,X^6 - 
    359890136718750000\,v_3\,X^6 + 
    4648425269468060592\,X^7 \\&- 
    5586218261718750000\,v_1\,X^7 + 
    4686492919921875000\,v_1^2\,X^7 + 
    683459472656250000\,v_2\,X^7 + 
    1297485351562500000\,X^8 \\&- 
    3417297363281250000\,v_1\,X^8 + 
    1153564453125000000\,X^9).
\end{split}
\end{equation}
}

\section{Generalization to the hypersurfaces in weighted projective space}
\label{app:2}
Here, we write the generators of the amplitudes for the hypersurfaces in 
weighted projective spaces $k=6,8,10$ in the notation of \cite{Klemm:1993tx}.
The generators $A_p,B_p,\ (p=1,2,3,\dots)$ are defined as the same way as 
eqs.\eqref{ABCX}. We also define $C=C_{\psi\psi\psi}\psi^3$ as in eq.\eqref{ABCX}.
On the other hand, $X$ is defined as
\begin{align}
 X=\frac{1}{1-\psi^k}.
\end{align}
The derivatives of these things are written as
\begin{align}
 \psi\del_{\psi}A_{p}=A_{p+1}-A A_{p},\quad 
 \psi\del_{\psi}A_{p}=A_{p+1}-A A_{p},\quad \psi\del_{\psi}X=kX(X-1),\quad
 \psi\del_{\psi}C=kXC.
\end{align}
The relation between generators are modified as follows. Eq.\eqref{br} is 
modified as
\begin{align}
 k=6,\qquad & B_{4}=12 X B_{3}-49X B_{2}+78 X B -40 X,\\
 k=8,\qquad & B_{4}=16 X B_{3}-86X B_{2}+176 X B -105 X,\\
 k=10,\qquad & B_{4}=20 X B_{3}-130X B_{2}+300 X B -189 X,
\end{align}
Eq.\eqref{ar} is  modified as
\begin{align}
 &A_2=-4B_2-2AB-2B+2B^2-2A+ 2 k XB + k XA-1-r_k X,\\
 &r_6=7,\qquad r_8=14,\qquad r_{10}=20.
\end{align}
The $u,v_1,v_2,v_3$ variables in eq.\eqref{uvsystem} are introduced as
\begin{equation}
 \begin{split}\label{uvsystem-apdx}
 &B=u,\qquad A=v_1-1-2u,\qquad
 B_2=v_2+uv_1,\\& B_3=v_3- u v_2+kX u v_1-(r_k+k)u X.
\end{split}
\end{equation}
The partition function $P_g:=C^{g-1}F_{g}$ can be written as a degree $(3g-3)$ 
inhomogeneous polynomial of $v_1,v_2,v_3,X$. For example, $P_2$ of $k=6$ 
hypersurface becomes
\begin{equation}
\begin{split}
P_2=& \frac{459}{20} - \frac{441}{8}v_1 + 
  \frac{21}{4}v_1^2 - 
  \frac{5}{24}v_1^3 - 
  \frac{357}{4}v_2 + 
  \frac{17}{4}v_1v_2 + 
  \frac{323}{8}v_3 - \frac{13873}{48} X- 
  7 v_1X +  v_1^2 X \\&- 
  \frac{493}{2}v_2 X + \frac{491}{240}X^2 - 
  \frac{13}{8}v_1 X^2 + \frac{9}{10}X^3.
\end{split}
\end{equation}
\providecommand{\href}[2]{#2}\begingroup\raggedright\endgroup

\end{document}